# Ultrafast, highly-sensitive infrared photodetectors based on two-dimensional oxyselenide crystals

**Authors:** Jianbo Yin[1†], Zhenjun Tan[1,2†], Hao Hong[3†], Jinxiong Wu[1†], Hongtao Yuan[4], Yujing Liu[1], Cheng Chen[5], Congwei Tan[1], Fengrui Yao[3], Yulin Chen[5], Zhongfan Liu[1,2], Kaihui Liu[2,3*], Hailin Peng[1,2*]

**Affiliations:**

[1] Center for Nanochemistry, Beijing Science and Engineering Centre for Nanocarbons, Beijing National Laboratory for Molecular Sciences, College of Chemistry and Molecular Engineering, Peking University, Beijing 100871, P. R. China

[2] Academy for Advanced Interdisciplinary Studies, Peking University, Beijing 100871, P. R. China

[3] State Key Laboratory for Mesoscopic Physics, School of Physics, Peking University, Beijing 100871, China

[4] National Laboratory of Solid-State Microstructures, College of Engineering and Applied Sciences, and Collaborative Innovation Center of Advanced Microstructures, Nanjing University, Nanjing 210093, P. R. China.

[5] Clarendon Laboratory, Department of Physics, University of Oxford, Parks Road, Oxford, OX1 3PU, UK

† These authors contributed equally to this work
*Correspondence: hlpeng@pku.edu.cn; khliu@pku.edu.cn

**Infrared detection and sensing is deeply embedded in modern technology and human society and its development has always been benefitting from the discovery of new photoelectric response materials[1]. The rise of two-dimensional (2D) materials[2], thanks to their distinct electronic structure, extreme dimensional confinement and strong light-matter interactions, provides new material platform for next-generation infrared photodetection[3-7]. Ideal infrared detectors should have fast respond, high sensitivity and air-stability, which is rare to meet at the same time for all existing 2D materials, either graphene, transition metal dichalcogenide or black phosphorous. Herein we demonstrate a new infrared photodetector based on 2D $Bi_2O_2Se$ crystals, whose main characteristics are superb in the whole 2D family: high sensitivity of ~65 A/W at 1200 nm and ultrafast intrinsic photoresponse of ~1 ps at room temperature. Such great performance is attributed to the suitable electronic bandgap and**

**high carrier mobility of 2D oxyselenide. With additional merits of mass production, excellent stability and flexibility, 2D oxyselenide detectors should open new avenues in highly-sensitive, high-speed, low-cost, flexible infrared photodetection and imaging.**

The discovery of new type of materials is the main driving force for the development of infrared (IR) photodetection beyond silicon's detection wavelength limit (~1100 nm). After century's efforts, different bulk materials with high quantum efficiency and tunable band gaps in the IR region, such as III-V and II-VI compounds of HgCdTe, InGaAs, InSb, GaAs/AlGaAs quantum wells, and InAs/GaSb super-lattices, have been discovered[1]. Currently, an important trend in IR detection is the combination of IR sensing materials with silicon readout circuit, enabling larger number of pixels, higher frame rates and more complicated on-chip signal-processing functions. To this end, great efforts have been devoted to finding new IR-sensitive materials with compatibility to silicon readout circuits, such as platinum silicide (PtSi) [8], black silicon[9], and quantum dots[10,11]. However, the sensitivity and response speed of these materials still have lots of room for the improvement.

The recent rise of 2D layered materials has opened up new possibilities for high-performing photodetection, thanks to their extreme dimensional confinement in the thickness and strong light-matter interactions in 2D plane[2-7]. In addition, 2D layered materials have excellent mechanical properties and dangling-bond-free interlayers, which allow for an easy processing of atomically thin layers into focal plan arrays (FPA) and compatibility with readout circuits[12,13]. However, 2D layered materials has not yet shown high sensitivity and fast response in detecting IR light. For example, graphene shows high-speed photoresponse[14] but very low sensitivity less than tens of mA/W[5]. Transition metal dichalcogenides (TMDs) usually have too large band gaps to detect IR light. Few-layer black phosphorus films exhibit promising infrared photodetection with high gain due to the fast carrier dynamics arising from its substantial mobility and moderate bandgap[15-17]. However, its environmental instability and incompatibility with large-scale fabrication processes have hindered its potential applications[18]. Therefore, identifying new air-stable 2D layered materials for highly-sensitive and high-speed IR detection is highly motivated.

Herein we report high-performing IR photodetectors based on a new air-stable 2D oxyselenide crystals at room temperature. The prototype devices of 2D $Bi_2O_2Se$ demonstrate a very high sensitivity of ~65 A/W at 1200 nm and an ultrafast intrinsic photoresponse of ~1 ps, which should originate from $Bi_2O_2Se$ crystal's appropriate band gap and ultrahigh carrier mobility. In combination with excellent flexibility (strain up to ~1%), high stability (months in air) and the capability of mass production, 2D $Bi_2O_2Se$ detectors hold promise in low-cost infrared imaging, high-speed photodetection and flexible biosensor operating at room temperature.

Very recently, $Bi_2O_2Se$ emerged as a promising new 2D layered material with excellent air stability and high-mobility semiconducting behavior[19]. As shown Fig. 1a, $Bi_2O_2Se$ is a layered oxyselenide that consists of alternatively stacked $Bi_2O_2$ and Se layers with a layer thickness of 0.61 nm. The unique lattice of $Bi_2O_2Se$ gives rise to an appropriate band gap of ~0.8 eV and relatively small electron effective mass of ~0.14 $m_0$, which are revealed by both theoretical calculations[19] and angle-resolved photoemission spectroscopy (ARPES, materials and methods) as shown in Fig. 1b. The layered nature also makes it ideal for fabricating electronic devices down to few atomic layers, which show high Hall mobility (~28900 $cm^2/V \cdot s$ at 1.9 K and ~450 $cm^2/V \cdot s$ at room temperature) and high current on/off ratio of >$10^6$ with almost ideal subthreshold swing (~65 mV/dec)[19]. Such high mobility facilitates the photocarrier extraction process, which, in combination with the appropriate band gap, underlies high sensitivity and fast response as introduced below, making 2D $Bi_2O_2Se$ a promising material for IR photodetection.

To understand the optoelectronic properties of 2D $Bi_2O_2Se$ crystals, we firstly synthesized individual $Bi_2O_2Se$ nanosheets with a domain size up to tens of μm and a thickness down to few layers on a mica substrate by chemical vapor deposition (CVD, see methods)[19,20], and then fabricated a photodetection device on it (Fig. 1c). We measured the photoresponse of 2D $Bi_2O_2Se$ crystal without applying external bias to guarantee the intrinsic photovoltaic response, and found a broadband response with the spectrum from visible to 1700 nm. In detail, we revealed photovoltage distribution by scanning a laser beam (150 μW, 1200 nm and ~1.5 μm in spot size) over the device (Fig. 1c), and recording the photovoltages with laser positions. As shown in Fig. 1d and 1e, prominent photovoltage signals were generated at two $Bi_2O_2Se$-metal junctions with opposite polarities, which results from photocarriers separation process—electron-hole pairs are separated to opposite directions by the symmetric build-in electrical fields at the two junctions. This phenomenon infers that the photocurrent generation is highly dependent on the photocarrier separation process, although other processes also play important roles, such as photocarrier excitation and relaxation.

To study the photocarrier excitation process of $Bi_2O_2Se$, we recorded photocurrents while changing the wavelengths of incident light and showed photovoltage spectrum in Fig. 1f and 1g. The spectrum extends to a remarkable value of ~1700 nm (~0.73 eV), which agrees in a reasonable accuracy with the band gap value (~0.8 eV) observed by ARPES and optical absorption in Fig. 1g. Such agreement confirms that the photocarrier excitation originates from interband transition. In addition, as the photon energy increases, the photocurrent increases in the same trend with the optical absorption, implying that photocurrent generation is also subject to the photocarrier generation process. This increasing trend is in agreement with the change of density of states (DOS) as shown in Fig. 1b—away from band gap, both conduction and valence bands have increasing DOS, favoring the interband transition.

As the photocarrier extraction is critical for the photocurrent generation, we can enhance the photocarrier extraction efficiency by deliberately increasing longitude electric field and creating a photoconductivity regime. This regime typically contains bias-dependent photocurrent and gain, which is usually larger than 1 and defined as the number of electrically circulated carriers per incident photon[21]. In this regime, photocurrent $I_{ph}$ is defined as difference of drain currents between dark and illuminated states (inset of Fig. 2a), and responsivity ($R$) is extracted by equation $R=I_{ph}/P$, where $P$ denotes incident power. To quantify the photosensitivity of 2D $Bi_2O_2Se$ devices, we summarized the responsivities at 1200 nm as scatter plot in Fig. 2a, and color chart in Fig. 2b with different biases. Both figures show a remarkable value of ~65 A/W at incident power of ~100 pW scale, implying the capability of $Bi_2O_2Se$ in detecting weak infrared signals, which is very important for practical applications. This excellent performance can be further evidenced by high sensitivities along the whole spectrum from visible to IR regime, such as ~5800 A/W at 532 nm, ~4 A/W at 1310 nm (Supplementary Fig. 1) and ~0.1 A/W responsivity at 1550 nm (Fig. 2a, 2c). In comparison with other 2D materials, which show low sensitivities in infrared spectral range, 2D $Bi_2O_2Se$ exhibits high sensitivity in an extremely broad spectral region of 300 ~1700 nm (Fig. 2c), much superior to other 2D materials such as graphene and TMD[22].

In addition to the sensitivity, the photoresponse speed is the other important criterion that directly determines the fundamental bandwidth of a photodetector[23]. We performed pump-probe experiments with ultrafast pulse laser (820 nm, 100 fs) to investigate the intrinsic photoelectric response speed limit of 2D $Bi_2O_2Se$ photodetector in the near-infrared (NIR) range (see supplementary note 1 and Supplementary Fig. 2 for details). As shown in a typical time-resolved photocurrent measurement of 2D $Bi_2O_2Se$ (Fig. 3a), the photocurrent reaches minimum when pump and probe pulses overlap in time domain. With time-delay increasing from zero, the photocurrent increases from minimum correspondingly. The dip in photocurrent curve directly reflects the sublinear power dependence of the photocurrent and the ability of distinguishing two time-nearby pulses[14,24-27]. With exponential fitting of the rising curve, we figured out the response time is ~1 ps at variable incident power (Fig. 3b), implying the detectable frequency limit of 2D $Bi_2O_2Se$ is as high as ~1 THz. Such response time is comparable to the reference graphene sample (~1.3 ps as shown by dark curve in Fig. 3a) and significantly shorter than any other 2D material (summarized in Supplementary Fig. 3). We believe that the ultrafast response of $Bi_2O_2Se$, in analog to 2D Dirac materials of graphene and 3D Dirac materials of $Cd_3As_2$[14,24,28], is originated from the ultrahigh electron mobility of $Bi_2O_2Se$, enabling the fast photocarrier extraction and relaxation. In contrast to the Dirac materials with zero bandgap, layered $Bi_2O_2Se$ is a 2D semiconductor with appreciable bandgap and therefore demonstrates a high free carrier generation efficiency with low energy loss. In such sense, 2D $Bi_2O_2Se$ should be a very promising material in high-speed infrared photodetection.

To better understand the superiority of 2D $Bi_2O_2Se$ in IR sensing applications, we designed and fabricated flexible 2D $Bi_2O_2Se$ photodetectors and their arrays on mica through a facile process (see Fig. 4a, 4e and supplementary note 2) and tested them in ambient air. The $Bi_2O_2Se$ photodetector arrays show consistent photoresponse when bending the substrate with strain of ~1% (inset of Fig. 4a), confirming that 2D $Bi_2O_2Se$ photodetector can work on flexible substrates. More importantly, they are quite robust when exposed in ambient air, manifesting very stable photoresponse within at least 5 weeks (Fig. 4b), which is critical for practical applications.

Furthermore, we demonstrate the imaging capability of $Bi_2O_2Se$ photodetectors by two approaches. Firstly, single-pixel photodetector is tested to image a specific structure by scanning infrared reflection from the sample as shown in Fig. 4c. A clear infrared image of the sample structure under 1150 nm illumination is shown in Fig. 4d, which confirms that a single photodetector of 2D $Bi_2O_2Se$ can function well in a near-infrared imaging system (see supplementary Fig. 5 for more details). Secondly, a 3×5 multi-pixel array as shown in Fig. 4e is examined by scanning the array over a steady laser beam. As several channels are pre- picked up in a parallel regime, the photocurrents from these pixels are read out and show corresponding shapes of "1", "2", "3", and "4" under different illuminations, with wavelengths of 1550 nm, 1310 nm, 1200 nm and 665 nm, respectively. These images from both single- and multi- pixels imply that $Bi_2O_2Se$ is capable to process into FPA for multi-spectrum imaging (from visible to near infrared). Such imaging capability, in combination with its superb photoresponse, as well as its chemical stability and easy accessibility, make 2D $Bi_2O_2Se$ a promising candidate for realizing ultrafast, highly-sensitive infrared photodetectors operating at room temperature.

**Methods**

**Growth and characterization of $Bi_2O_2Se$**

The $Bi_2O_2Se$ nanoplates were synthesized via a previously reported chemical vapor deposition (CVD) method. The as-synthesized samples were characterized by Olympus BX51 microscope. The ARPES measurements of $Bi_2O_2Se$ were carried out at beamline I05 of the Diamond Light 48 Source (DLS), with energy resolution of 20 meV and angle resolution of 0.2º.

**Device fabrication and measurement**

Bi$_2$O$_2$Se devices are made on mica after growth and on Si/SiO$_2$ (300 nm) after transfer with help of poly(methyl methacrylate). The Ti/Au (5/35 nm) electrodes were fabricated by electron-beam lithography and the following electron-beam evaporation. The photoelectrical measurements were performed by a scanning photocurrent microscopy. In the set-up, Supercontinuum Laser Sources (NKT Photonic) were used as laser sources. The chopper-modulated (~500 Hz) laser beams were focused to ~2 μm on the device using ~50 objective and the short-circuit photocurrents were then measured by pre-amplifier and lock-in amplifier, while the photovoltages were directly measured by lock-in amplifier. When scanning the laser spot over the device, the induced photocurrents (photovoltage) and beam positions were recorded and displayed simultaneously with the assistance of a computer, which communicated with lock-in amplifier and motorized stage (with device on it). In the photoconductivity measurement, Keithley 2400 was used to supply the source-drain bias and at the same time measure the current.

For time-resolved measurement, two cross-polarized (minimize interference near zero delay) pulsed beams (both with pulse width of ~100 fs and wavelength of ~820 nm. The laser is generated from a Ti: sapphire 80 MHz oscillator) were focused on Bi$_2$O$_2$Se or graphene samples. One pulsed beam was chopped at frequency of 1500 Hz to generate modulated photovoltage in the device circuit. This photovoltage was sensitive to the presence of the second beam pulses, which was temporally delayed by $\Delta t$ through a motorized delay line. Thus the photovoltage could be researched as function of the delay time. The sublinear photovoltage-power measurement was conducted without the presence of the second beam. The response time is defined as time constant $\tau$ in exponential fitting of $-A * \exp * (\frac{-\Delta t}{\tau})$.

**Acknowledgments:**
We acknowledge financial support from the National Basic Research Program of China (Nos. 2014CB932500 and 2016YFA0200101) and the National Natural Science Foundation of China (Nos. 21733001, 21525310 and 51522201).

**Author contributions**
H.P., J.Y., and K.L. conceived and designed the experiments. J.W. and C.T. performed the synthesis, and optical absorption measurement. J.W., Z.T., Y.L. and J.Y. fabricated devices. C.C. and Y.L.C. conducted ARPES measurements and analysis. H.H., K.L., Z.T., F.Y. and J.Y. performed time-resolved photocurrent measurement. Z.T and J.Y. performed single-pixel infrared imaging measurement, while J.Y. performed all the other optoelectronic measurement. J.Y., H.P. and K.L. wrote the paper with input from the other authors. H.P. supervised the project. Z.L., H.Y. and all the other authors contributed to the scientific discussions.

**Additional information**

Supplementary information is available in the online version of the paper. Reprints and permissions information is available online at www.nature.com/reprints. Correspondence and requests for materials should be addressed to Hailin Peng and Kaihui Liu.

**Competing financial interests**

The authors declare no competing financial interests.

**Figures and captions**

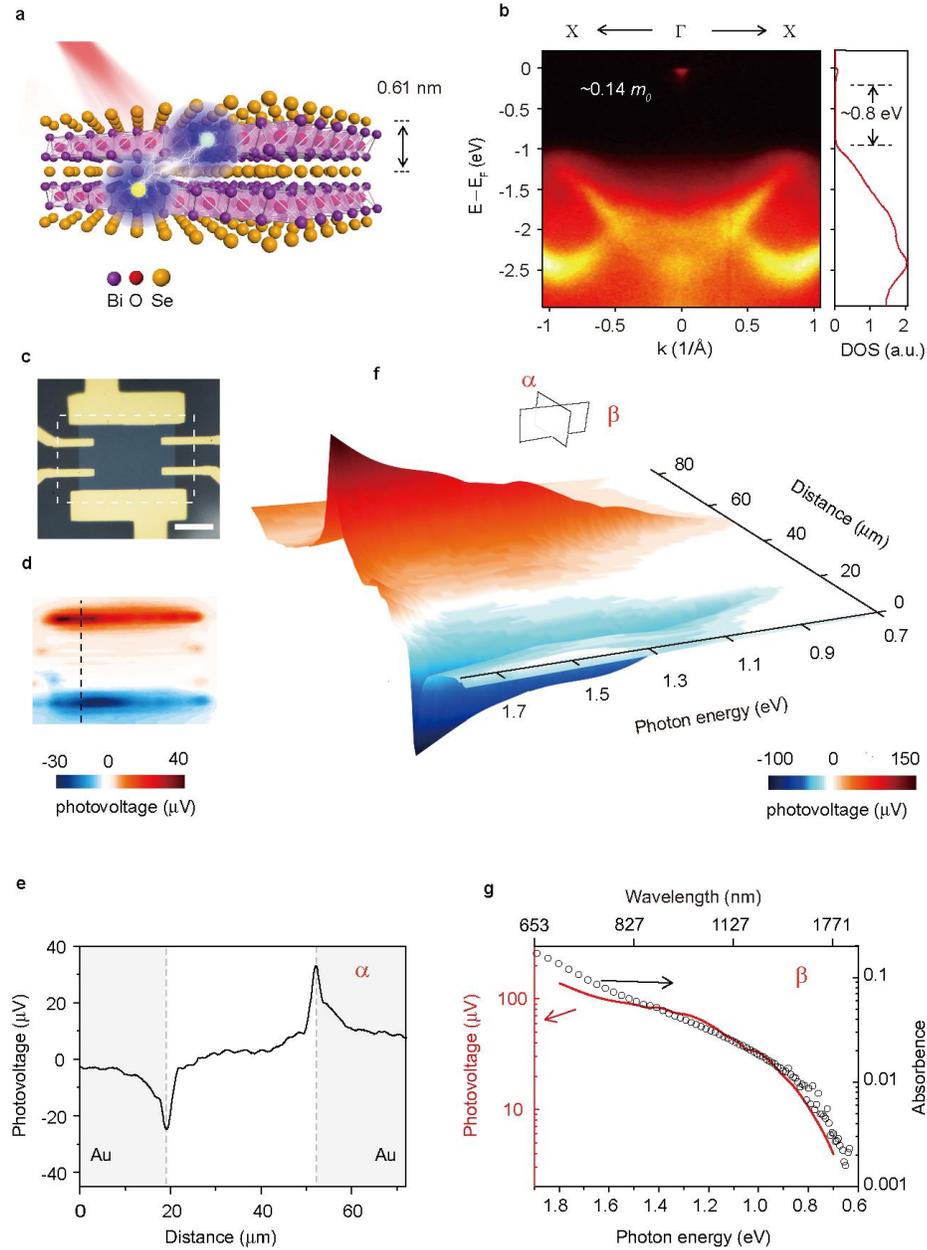

**Figure 1 | Photodetectors based on Bi$_2$O$_2$Se layered materials.** (**a**) Illustration of photodetector based on Bi$_2$O$_2$Se crystal, with schematic crystal structure of alternatively-stacked (Bi$_2$O$_2$)$_n$ and Se$_n$ layers. The layer thickness is 0.61 nm. (**b**) Electronic structure of Bi$_2$O$_2$Se observed by ARPES, which shows an indirect band gap of ~0.8 eV and small effective mass of ~0.14 $m_0$. The directions of ARPES mapping are along X-Γ-X directions. The right panel is the corresponding density of states (DOS). (**c**) Optical image of Bi$_2$O$_2$Se device, scale bar 20 μm. (**d**) Scanning photovoltage image of the dotted rectangle area of device in (**c**). The 1200 nm laser with power of 150 μW is focused and scanned on the device, while the net photovoltages are recorded as function of laser positions. The photovoltages are measured without external bias. (**e**) Line-scanning of

photovoltage along dotted line in (**d**). The curve corresponds to the *α*-plane cut of (**f**). The two peaks correspond with $Bi_2O_2Se$-metal junctions. (**f**) Spectrogram composed of photovoltage line scans at different incident photon energies (wavelength of incident light). The incident photon numbers are kept still during the measurement, and the incident power is 100 μW at 1200 nm. (**g**) Red solid line is the photovoltages as function of incident photon energies, which corresponds to *β*-plane cut of the Spectrogram. Dashed line is the absorbance of $Bi_2O_2Se$ film with ~10 nm in thickness.

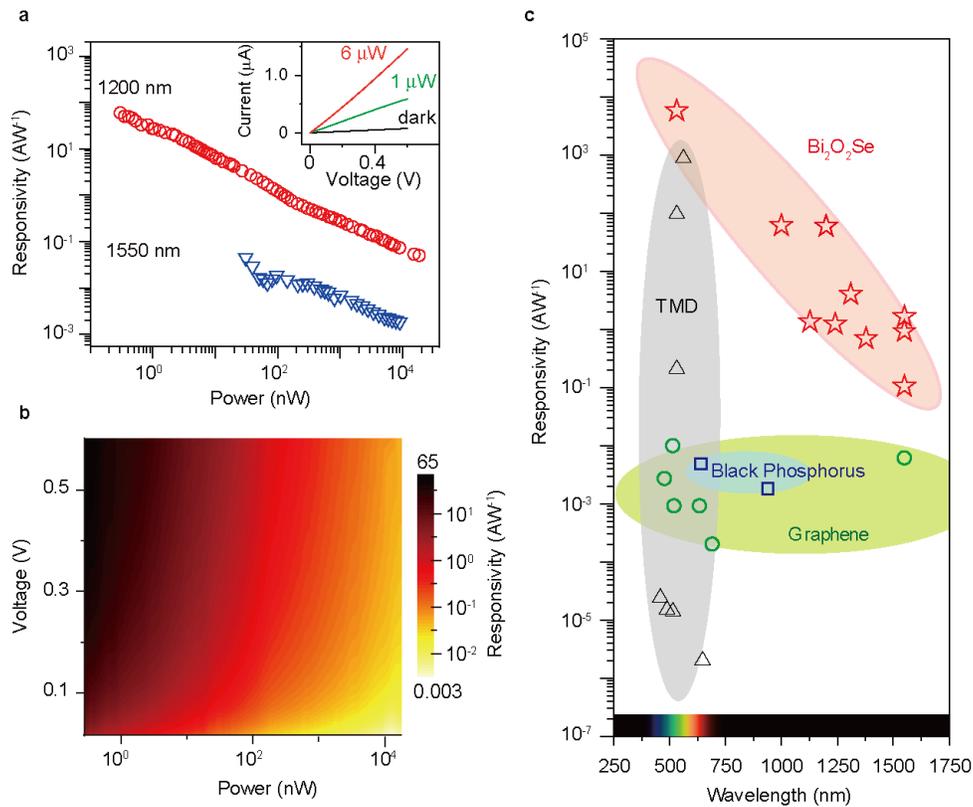

**Figure 2 | High photoresponsivity of $Bi_2O_2Se$ photodetector.** (**a**) Photoresponsivity under 0.6 V bias at wavelengths of 1200 nm and 1500 nm. (**b**) Dependence of photoresponsivity with incident power and voltage bias at 1200 nm wavelength. (**c**) Comparison of photodetectors based on $Bi_2O_2Se$, graphene, black phosphorus, and transition metal dichalcogenides (TMDs) *(14)*. Note that the data only includes the photocurrent generation through excitation of interband transitions without further treatment such as adding waveguide structure or plasmonic structure.

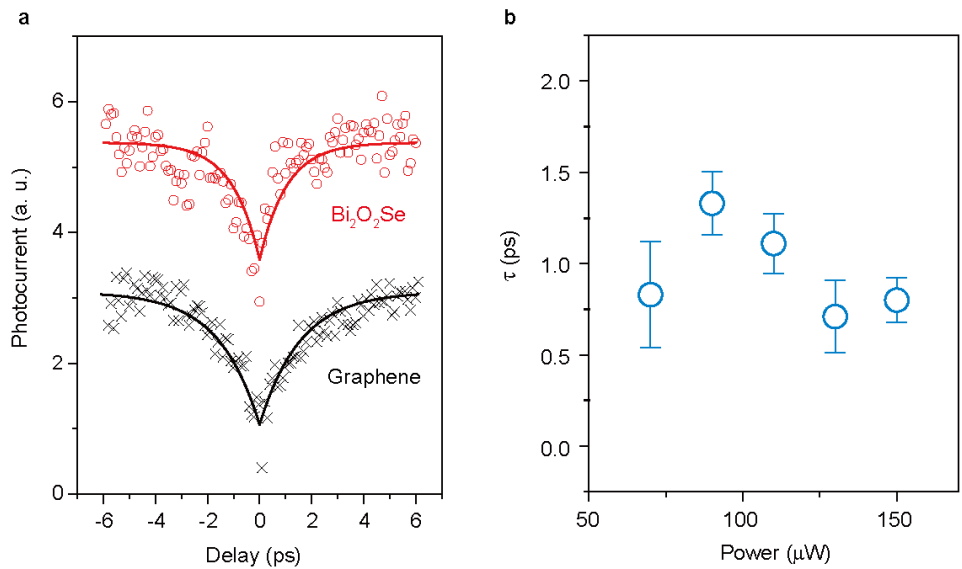

**Figure 3 | Ultrafast photoresponse of Bi$_2$O$_2$Se photodetector.** (**a**) Photocurrent as function of delay time between two ultrafast pulses (~100 fs). The red circles and black crosses are data of Bi$_2$O$_2$Se and graphene photodetector, while solid lines are exponential fitting. When delay time is too short, the excitation of the first pulse does not have enough time to turn into current before it meets the excitation of the second pulse, which induces sublinearity in photocurrent and reaches a dip at zero delay. When delay time is long enough, the two pulse excitations behave independently and result in largest voltage, which is used for normalization value in this chart. (**b**) Response time of Bi$_2$O$_2$Se photodetector measured at different incident power.

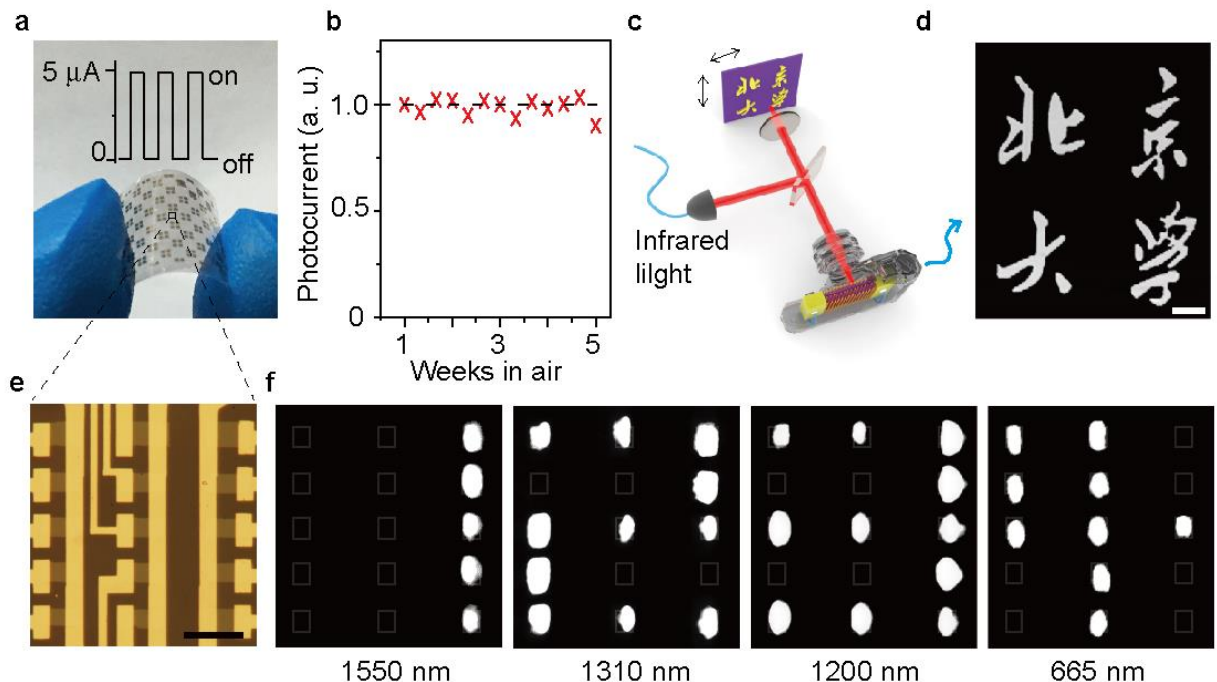

**Figure 4 | Flexible Bi$_2$O$_2$Se photodetector arrays.** (**a**) Photograph of Bi$_2$O$_2$Se photodetectors and arrays on mica, Inset shows photoresponse of one typical photodetector when bending the substrate with strain up to 1%. The current show robust increase when illumination of 1200 nm ~100 μW is on. (**b**) Bi$_2$O$_2$Se photodetector maintains stable photoresponse in air for at least 5 weeks. (**c**) Schematic of the single-pixel imaging process. The Bi$_2$O$_2$Se photodetector quantify the infrared reflection of the sample structure by measuring corresponding photocurrent. When the structure is scanned, the detector records the magnitudes of reflection signals and process them into image. (**d**) Infrared images taken by single-pixel photodetector of Bi$_2$O$_2$Se under excitation at λ = 1150 nm. Scale bar, 100 μm (**e**) Optical image of 3×5 multi-pixel array of Bi$_2$O$_2$Se photodetectors. Scale bar, 10 μm. (**f**) The array's photocurrent mappings are achieved by recording the photocurrents while scanning the array over a pre-focused laser beam with spot size of 1~2 μm. By deliberately choosing several channels in a parallel regime, photocurrents from the chosen pixels are recorded and shown in the photocurrent mapping. Here, the resulting photocurrent images with shape of "1", "2", "3" and "4" are taken under illumination of 1550 nm, 1310 nm, 1200 nm and 665 nm light, respectively.